\documentclass[11pt, letter]{article}
\usepackage{rubaton-light}
\usepackage[english]{babel}
\usepackage{blindtext}\blindmathtrue
\usepackage{relsize}
\usepackage{avant} 
\usepackage{amsmath,epsf,amssymb,latexsym,amsthm,setspace,array,pifont,color,hyperref,amsfonts,dsfont,cancel,braket,parskip}
\usepackage[none]{hyphenat} 
\usepackage{graphicx, feynmp, feynmp-auto} 
\usepackage{feynman,slashed, feyn, comment, todonotes}
\graphicspath{{Pictures/}} 

\newcommand{\cl}[1]{\mathcal{#1}}

\renewcommand{\to}{\longrightarrow}

\newcommand{\Trm}[1]{\text{Tr$_-$}\left(#1\right)}
\newcommand{\Trp}[1]{\text{Tr$_+$}\left(#1\right)}
\newcommand{\Trpm}[1]{\text{Tr$_\pm$}\left(#1\right)}
\def\spvec#1{\left(\vcenter{\halign{\hfil$##$\hfil\cr \spvecA#1;;}}\right)}
\def\spvecA#1;{\if;#1;\else #1\cr \expandafter \spvecA \fi}

\title{Comments on scattering in massive gravity, vDVZ and BCFW}
\author{Nathan Moynihan and Jeff Murugan}

\affiliation{The Laboratory for Quantum Gravity \& Strings,\\ 
	Department of Mathematics \& Applied Mathematics,\\
	University Of Cape Town,\\
	Private Bag, Rondebosch, 7701, South Africa}
\emailAdd{nathantmoynihan@gmail.com}
\emailAdd{jeff.murugan@uct.ac.za}

\abstract{
Armed with the latest technology in the computation of scattering amplitudes involving massive particles of any spin, we revisit the van Dam-Veltman-Zakharov (vDVZ) discontinuity of massive gravity and show how it may be understood in terms of the Britto-Cachazo-Feng-Witten (BCFW) relations.}

\begin{document}
	
\maketitle
\begin{fmffile}{diagrams}
	\section{Introduction}
	The idea that the graviton, the quantum of gravity, may have a small but non-vanishing
	mass is one that has been around since Fierz and Pauli's original work on massive 
	spin-2 field theory. Phenomenologically, there is much appeal to a theory in which
	General Relativity is modified in this way at large distances, not the least of which is 
	a possible explanation of the current acceleration of the Universe that does not 
	invoke any dark energy. Unfortunately, massive gravity also suffers from a range of 
	pathologies that, at least historically, have severely constrained its viability. These 
	include the presence of the Boulware-Deser ghost and a discontinuity with 
	General Relativity (GR) as the graviton mass is sent to zero. While we will have 
	nothing to contribute to the discussion of ghosts, it will be the so-called vDVZ 
	discontinuity \cite{1970JETPL..12..312Z, vanDam:1970vg} that will form the basis for this article.   
        
	The inability of the massive theory to smoothly reduce to GR in the limit that 
	the mass of the graviton is taken to zero famously manifests in a gravitational 
	lensing angle only three quarters of the observed value. Physically, this is understood
	by observing that a massive spin-2 field propagated three additional degrees of freedom
	than its massless counterpart. These are repackaged as a vector and a scalar, and it is found that the scalar
	couples to the trace of the stress-energy tensor of any matter coupled to the massive gravity, 
	providing an additional force. In order to reconcile this with the classical Newtonian
	potential, the gravitational coupling must be rescaled to three quarters its value in the
	Einstein theory. However, since the gravitational lensing of light is blind to the scalar (its stress energy tensor being traceless), this 
	results in a proportionally smaller lensing angle than that computed in GR. In the interests 
	of pedagogy, let's unpack the details of this argument. 
	
	Giving the graviton mass breaks the full diffeomorphism invariance of GR. This can be reintroduced via the
    St\"uckelberg procedure, but at the expense of the introduction of several new fields. Starting from an action with explicitly broken diffeomorphism symmetry, 
     and involving only a single dynamical field $h_{\mu\nu}$, 
	\begin{equation}\label{key}
		S[h] = \int d^4x~\sqrt{-g}\left(\frac{2}{\kappa^2}R -\frac{m_h^2}
		{2}\left(h_{\mu\nu}h^{\mu\nu} - h^2\right) \right)\,,
	\end{equation} 
	we then demand that diffeomorphism symmetry is restored by transforming 
	$h_{\mu\nu}$ by a St\"uckelberg field (scaled by the graviton mass for convenience) 
	that encodes the transformation
	\begin{equation}\label{key}
		\delta h_{\mu\nu} = \frac{1}{m_h}\left(\partial_\mu A_\nu + \partial_\nu 
		A_\mu\right)\,.
	\end{equation}
	The result is the new action
	\begin{equation}\label{key}
		S[A,h] = \int d^4x~\sqrt{-g}\left(\frac{2}{\kappa^2}R -\frac{m_h^2}
		{2}\left(h_{\mu\nu}h^{\mu\nu} - h^2\right) - \frac12 F_{\mu\nu}F^{\mu\nu}
		 -2m_h(h_{\mu\nu}\partial^\mu A^\nu -  h\partial_\mu A^\mu)\right)\,.
	\end{equation}
	Subsequently, demanding the gauge invariance of the vector field requires the introduction of
	another St\"uckelberg field, this time a scalar, via the transformation
	\begin{equation}
		\delta A_\mu = \frac{1}{m_h}\partial_\mu\pi\,,
	\end{equation}
     and results in the action
	\begin{eqnarray}
		S[A,h,\pi] = \int d^4x~\sqrt{-g}\bigg(&\frac{2}{\kappa^2}R -\frac{m_h^2}
		{2}\left(h_{\mu\nu}h^{\mu\nu} - h^2\right) - \frac12 F_{\mu\nu}F^{\mu\nu} 
		-2m_h(h_{\mu\nu}\partial^\mu A^\nu -  h\partial_\mu A^\mu)\nonumber \\
		&- 2(h_{\mu\nu}\partial^\mu \partial^\nu\pi -  h\partial^2\pi)\bigg)\,.
	\end{eqnarray}
	The massive graviton can be coupled to a source $T^{\mu\nu}$ through a source term 
	$\kappa h_{\mu\nu}T^{\mu\nu}$, whose variation (after integration by parts) is
	\begin{align}\label{key}
		\delta\left(h_{\mu\nu}T^{\mu\nu}\right) 
		=  \frac{2\pi}{m_h^2}\partial_\mu \partial_\nu T^{\mu\nu} - \frac{2A_\nu}{m_h}
		\partial_\mu T^{\mu\nu}\,.
	\end{align}
     Assuming stress-energy conservation (i.e. $\partial_\nu 
     T^{\mu\nu} = 0$), this variation is zero, resulting in the diffeomorphism invariant 
     sourced theory with action  
	\begin{align}
		S[A,h,\pi] = \int d^4x~\sqrt{-g}\bigg(&\frac{2}{\kappa^2}R -\frac{m_h^2}
		{2}\left(h_{\mu\nu}h^{\mu\nu} - h^2\right) - \frac12 F_{\mu\nu}F^{\mu\nu} 
		-2m_h(h_{\mu\nu}\partial^\mu A^\nu -  h\partial_\mu A^\mu)\nonumber \\
		&- 2(h_{\mu\nu}\partial^\mu \partial^\nu\pi -  h\partial^2\pi) +\kappa h_{\mu\nu}
		T^{\mu\nu}\bigg)\,.
	\end{align}
	Currently, the $h_{\mu\nu}$ tensor still represents all 5 modes of the graviton, but it can be explicitly 
	decomposed 
	it into the spin-2 and spin-0 modes in an effort to understand what happens to the kinetically mixed scalar-tensor
    modes\footnote{We will not consider the spin one mode, since the spin one Stuckelberg field is free}. 
    To this end, let's make a canonical transformation of the form
	\begin{equation}\label{key}
      	h_{\mu\nu} = \bar{h}_{\mu\nu} + \chi\eta_{\mu\nu}
	\end{equation}
     Where $\bar{h}$ is the tensor mode and $\chi$ the scalar. To linear order, the massless spin-2 part transforms as
	\begin{equation}\label{key}
		\int d^4x~\sqrt{-g}\frac{2}{\kappa^2}R \longrightarrow \int d^4x~\sqrt{-g}\frac{2}{\kappa^2}\bar{R} 
		+ 2\left(\partial_\mu\chi\partial^\mu\bar{h} - \partial_\mu\chi\partial_\nu\bar{h}^{\mu\nu} 
		+ \frac{3}{2}\partial_\mu\chi\partial^\mu\chi\right)\,,
	\end{equation}
	so that defining $\chi = \pi$ and with a little more manipulation, the action becomes 
	\begin{align}\label{key}
		S[A,\bar{h},\chi] = \int d^4x~\sqrt{-g}\bigg(&\frac{2}{\kappa^2}\bar{R} -\frac{m_h^2}{2}
		\left(\bar{h}_{\mu\nu}\bar{h}^{\mu\nu} - \bar{h}^2\right) - \frac12 F_{\mu\nu}F^{\mu\nu} 
		-2m_h(\bar{h}_{\mu\nu}\partial^\mu A^\nu -  \bar{h}\partial_\mu A^\mu)\nonumber \\
		&+\kappa \bar{h}_{\mu\nu}T^{\mu\nu} + 3\chi(\partial^2 + 2m_h^2)\chi + 3(m_h^2\bar{h}\chi 
		+ 2m\chi\partial_\mu A^\mu) + \kappa\chi T\bigg)\,.
	\end{align}
	Now that all of the degrees of freedom are accounted for, we can send $m_h\longrightarrow 0$ smoothly. 
	In this limit, we find\footnote{For the terms $\propto \frac1m \partial_\mu T^{\mu\nu}$, which we have ignored, we have assumed that $\partial_\mu T^{\mu\nu} \longrightarrow 0$ faster than $m\longrightarrow 0$.}
	\begin{align}\label{key}
		S[A,\bar{h},\chi] 
		\to \int d^4x~\sqrt{-g}\bigg(\frac{2}{\kappa^2}\bar{R} - \frac12 F_{\mu\nu}F^{\mu\nu} 
		+\kappa \bar{h}_{\mu\nu}T^{\mu\nu} - \frac12\partial_\mu\chi\partial^\mu\chi 
		+ \sqrt{\frac16}\kappa\chi T\bigg)\,.
     \end{align}
We recognise this as a theory containing an interacting massless scalar field (with a canonical kinetic term), an interacting massless spin-2 graviton and a free spin-1 field. Importantly, the scalar graviton couples to the trace of the stress energy tensor, so that any matter with a traceless stress energy tensor will not feel the effects of the scalar graviton. 
Of course, the canonical example of such matter is the photon of the electromagnetic interaction and a direct consequence of the above is that, if massive gravity and GR are to agree on their nonrelativistic Newtonian potential, then the bending angle of gravitationally lensed light must be qualitatively different between the two. Viewed as a scattering process, gravitational lensing corresponds to the Feynman diagram,\bigskip
		\begin{equation}\label{fdiags1}
		\begin{gathered}
		\begin{fmfgraph*}(120,80)
		\fmfleft{i1,i2}
		\fmfright{o1,o2}
		\fmf{photon,label=$1^{\mp1}$,label.side=left}{i1,v1}
		\fmf{plain,label=$4^0$,label.side=left}{i2,v1}
		\fmf{photon,label=$2^{\pm1}$,label.side=left}{v1,o1}
		\fmf{plain,label=$3^0$,label.side=left}{v1,o2}
		\fmfblob{.30w}{v1}
		\end{fmfgraph*}
		\end{gathered}
		~~~=~~~\mathlarger{\mathlarger{\mathlarger{\mathlarger{‎‎\sum}}}}_{s}
		\begin{gathered}
		\begin{fmfgraph*}(120,80)
		\fmfleft{i1,i2}
		\fmfright{o1,o2}
		\fmf{photon,label=$1^{\mp1}$,label.side=left}{i1,v1}
		\fmf{fermion,label=$4^0$,label.side=left}{i2,v2}
		\fmf{dbl_wiggly}{v1,v2}
		\fmf{photon,label=$2^{\pm1}$,label.side=left}{v1,o1}
		\fmf{fermion,label=$3^0$,label.side=left}{v2,o2}
		\fmflabel{$_{-s}$}{v1}
		\fmflabel{$_{s}$}{v2}\,,
		\end{fmfgraph*}
		\end{gathered}
		\end{equation}
		where the sum is taken over the two tensor, two vector and one scalar polarization modes of the 
		massive graviton. In this note, we will consider the above scattering process using the BCFW recursion
		relations, without resorting to the necessity of Lagrangians, polarization vectors or gauges and 
		show that the vDVZ discontinuity exists at the level of the amplitudes, regardless of the underlying 
		off-shell theory.

\section{Massive 3-point functions from SU(2)}
We will begin by reviewing the methods required to derive three point amplitudes from the little group in the case of massive particles. The usual approach to describing massive particles within the spinor helicity formalism is to make a decomposition of the form \cite{Boels:2009bv}
\begin{equation}\label{massivedecomp}
   P_\mu = k_\mu + \frac{P^2}{2q\cdot k}q_\mu\,,
\end{equation}
where $P$ is a massive vector and $k$ and $q$ are lightlike vectors. While $k_\mu$ is a unique lightlike vector, $q_\mu$ can be freely chosen provided $q\cdot k \neq 0$ and $q\cdot P \neq 0$. In effect, this gives a representation of massive vectors as massless ones, which can then be represented by spinors. Schematically we can write this as
\begin{equation}\label{massivespinordecomp}
   P_{massive} = \lambda\tilde{\lambda} + \alpha\eta\tilde{\eta}\,.
\end{equation}
It will prove more convenient, however, to follow the methods presented in \cite{Arkani-Hamed:2017jhn}. In this formalism we would instead demand a decomposition
\begin{equation}\label{key}
   P_{\alpha\dot{\alpha}} = \lambda_\alpha^I\tilde{\lambda}_{\dot{\alpha}I}\,.
\end{equation}
Here $I=1,2$ is an $SL(2)$ index that transforms under the $SU(2)$ subgroup for real, Lorentzian momenta as
\begin{equation}\label{key}
   \lambda^I_\alpha\longrightarrow W^{I}_J\lambda^J_\alpha\,.
\end{equation}
These indices are raised and lowered with $\epsilon^{IJ}$. In this new language, the equivalent of the Dirac equation
reads
\begin{equation}\label{key}
   P_{\alpha\dot{\alpha}}\lambda^{\alpha I} = -m\tilde{\lambda}_{\dot{\alpha}}^I,~~~~~~~~~P_{\alpha\dot{\alpha}}   
   \tilde{\lambda}^{\dot{\alpha} I} = m\lambda_{\alpha}^I\,.
\end{equation}
Conversion between dotted and undotted indices is facilitated by the operator
\begin{equation}\label{key}
   (J_i)^\alpha_{~\dot{\alpha}} = \frac{(P_i)^\alpha_{~\dot{\alpha}}}{m}\,.
\end{equation}
Key to this formulation of the problem is Wigner's ``little group" that governs the kinematics of particle scattering.
For massless particles, the kinematical on-shell constraints are trivialized through the introduction of little group-adapted vaiables like spinor-helicity, twistor or momentum-twistor variables. This is turn allows for one to side-step quantum fields and all their subtlties and work directly with the concept of a particle.  
Since the little group for massive particles is $SU(2)$, amplitudes must be constructed by working with objects that transform appropriately under $SU(2)$. Specifically, these are symmetric tensors with $2S$ indices, where $S$ is the magnitude of the total spin of the particle. In what follows, we will choose to express these amplutudes in a purely chiral basis, meaning that the constructed objects are indexed by $\alpha_1\alpha_2\cdots\alpha_{2S}$, using the operator we just defined.

Summarising the results of \cite{Arkani-Hamed:2017jhn}, a general strategy for constructing 3-point amplitudes is as follows:\begin{itemize}
	\item For each massless leg, assign a helicity $h_i$.
	\item For each massive leg of spin $S$, assign 2S spinor indices, $\alpha_1\alpha_2\cdots\alpha_{2S}$
	\item Using physical variables with spinor indices ($\lambda_{\alpha}, P_{\alpha\dot{\alpha}}$ etc) and the conversion operator defined above, construct a basis of $SL(2)$ 
	\item Write down every possible unique, maximally symmetric object with $2S$ indices in the newly constructed basis to get the \textbf{stripped amplitude} $$M_{\{\alpha_1\alpha_2\cdots\alpha_{2S_1}\}, \{\beta_1\beta_2\cdots\beta_{2S_2}\}, \{\gamma_1\gamma_2\cdots\gamma_{2S_3}\}}$$
	\item Contract each massive leg $i$ with $2S$ massive spinors $\lambda_i^I$ to find the final amplitude, which should now be labelled with helicities $h$ and $SL(2)$ indices $I,J,K...$.
	$$M^{I_1I_2...I_{2S},h_i,h_j...} = (\lambda_1)^{I_1\alpha_1}\cdots(\lambda_1)^{I_{2S_1}\alpha_{2S_1}}\cdots(\lambda_3)^{J_1\gamma_1}\cdots(\lambda_1)^{J_{2S_3}\gamma_{2S_3}}M_{\{\alpha_1\alpha_2\cdots\alpha_{2S_1}\}, \{\beta_1\beta_2\cdots\beta_{2S_2}\}, \{\gamma_1\gamma_2\cdots\gamma_{2S_3}\}}$$
\end{itemize}
In order to determine the amplitude for any one particular helicity configuration then, we simply project it out by contracting this stripped amplitude with the appropriate combination of chiral spinors and select the appropriate $SL(2)$ indices. This is essentially because the massive spinors can be expresses in a basis that is aligned and anti-aligned with the direction of the spinor, {\it i.e.}
\begin{equation}\label{key}
\lambda^I = \lambda \zeta^{-I} + \eta\zeta^{+I}
\end{equation}

Where $\zeta^{+I} = \spvec{1;0}$ and $\zeta^{+I} = \spvec{0;1}$

In this basis, the negative and positive helicity components are selected by $I = 1$ and $I = 2$ respectively.
 
For a massive particle with momentum $P_k$, we can choose the convention that $\braket{k\eta_k}[k\eta_k] = m_k^2$ and therefore that $[k\eta_k] = \braket{k\eta_k} = m_k$. For massless particles, contractions of like spinors are zero, $[ii] = \braket{ii} = 0$, but for massive particles (using \textbf{bold} notation) this is no longer true
\begin{equation}\label{key}
[\textbf{k}_I\textbf{k}_J] = \braket{\textbf{k}_I\textbf{k}_J} = \begin{cases} 
m & I>J \\
-m & I<J \\
0 & I=J 
\end{cases}
\end{equation}
We also note the useful identity $\bra{i}P_kP_k\ket{j} = -\bra{j}P_kP_k\ket{i} = m_k^2\braket{ij}$, which can easily be proved using the Schouten identity.
\subsection{High energy limit}
In what follows, we will need to sometimes take the high energy limit of particular massive amplitudes. Naively, it would seem like this should be implemented by sending $\eta \longrightarrow 0$ in eq. \ref{massivespinordecomp}. This is, however, too naive. In general such amplitudes contain terms of the form $\frac{\braket{\eta i}}{m}$ and so in the limit $\eta,m\longrightarrow 0$, are indeterminate. To circumvent this, a more sensible alternative is presented in \cite{Arkani-Hamed:2017jhn}, where 
\begin{equation}\label{helimit}
  \ket{\eta}\longrightarrow m\ket{\bar{\eta}},~~~~~~~|\eta]\longrightarrow m|\bar{\eta}]\,,
\end{equation} 
and, should either case arise explicitly, $\braket{\lambda\bar{\eta}}$ and $[\lambda\bar{\eta}]$ are set to unity, before taking the $m\longrightarrow 0$ limit.

\subsection{Calculation of 3-point amplitudes}
Now let's see how this works in detail by considering the 3-point function for two massive scalars and a spin-2 massive graviton. Since all the scattering particles are massive, the (chiral) $SL(2)$ space is spanned by the tensors\begin{equation}\label{key}
P_{1\alpha\dot{\beta}}(P_2)^{\dot{\beta}}_{~\beta} \equiv P_{\alpha\beta},~~~~~\varepsilon_{\alpha\beta}\,,
\end{equation}
which will form the basis for the amplitude. Moreover, since gravity couples universally with coupling $\sqrt{G} \sim \kappa$, and any 3-point function must have mass-dimension 1, the form of the amplitude can be read off from these building blocks essentially by dimensional analysis. The necessary maximally symmetric building blocks with $2S = 4$ indices constructed from these tensors are
\begin{equation}
  P_{\{\alpha_1\alpha_2}P_{\alpha_3\alpha_4\}} + P_{\{\alpha_1\alpha_2}\varepsilon_{\alpha_3\alpha_4\}} +   
  \varepsilon_{\{\alpha_1\alpha_2}\varepsilon_{\alpha_3\alpha_4\}}
\end{equation}
The associated stripped amplitude is then simply
\begin{equation}\label{key}
  M_{\{\alpha_1\alpha_2\alpha_3\alpha_4\}} = \kappa\left(P_{\{\alpha_1\alpha_2}P_{\alpha_3\alpha_4\}} 
  + P_{\{\alpha_1\alpha_2}\varepsilon_{\alpha_3\alpha_4\}} 
  + \varepsilon_{\{\alpha_1\alpha_2}\varepsilon_{\alpha_3\alpha_4\}}\right)\,.
\end{equation}

At this point, we could well express each amplitude with all of the $IJKL$ indices lavishly decorating each piece, but since symmeterisation of the spinor indices translates directly into symmeterisation of the $SL(2)$ indices, this will not be necessary. We will adopt the notation set out in \cite{Arkani-Hamed:2017jhn}, and represent massive spinors in \textbf{bold}. We will also suppress the $SL(2)$ indices, in the knowledge that they can always be reinstated in an unambiguous way (they are simply attached to all particles with spin and maximally symmetrized). Weighting each term by the graviton mass to give the correct mass dimension, the full amplitude is then given by
\begin{align}\label{mscalaramp}
M^{\left\{IJKL\right\}} &= M_{\{\alpha_1\alpha_2\alpha_3\alpha_4\}}\lambda_3^{I\alpha_1}\lambda_3^{J\alpha_2}\lambda_3^{K\alpha_3}\lambda_3^{L\alpha_4}\nonumber\\
&= \kappa\frac{([\textbf{12}]\braket{\textbf{13}}\braket{\textbf{23}})^2}{6m_h^4} + \kappa\frac{[\textbf{12}]\braket{\textbf{13}}\braket{\textbf{23}}\braket{\textbf{33}}}{4m_h^2} + \frac{\kappa}{2}\braket{\textbf{33}}^2\,,
\end{align}
where the numerical coefficients reflect the number of equivalent ways we can order $I,J,K,L$ in each term.

Unfortunately, this form of the amplitude does not lend itself to a direct application of the BCFW relations since we are unable to find a shift that is valid for all possible choices of helicity. Instead, we will need to extract the individual helicity components (in some limit where the helicity is well defined) and calculate the amplitudes for each helicity individually. This corresponds to making particular choices of $I,J,K,L$. It should be noted that this is entirely equivalent to contracting the stripped amplitude with an appropriate number of polarization vectors (or tensors). In the language of massive spinor-helicities, these are 

\begin{equation}\label{key}
\epsilon^-_{\alpha\beta} = \frac{\lambda_{\alpha}\lambda_{\beta}}{m},~~~~~		\epsilon^0_{\alpha\beta} = \frac{\lambda_{\alpha}\eta_{\beta} + \eta_{\alpha}\lambda_{\beta}}{2m},~~~~~\epsilon^+_{\alpha\beta} =\frac{\eta_{\alpha}\eta_{\beta}}{m}\,.
\end{equation}

That said, as card-carrying disciples of the ``on-shell" philosophy, we would prefer to work without the need for polarization vectors whatsoever. From the amplitude \ref{mscalaramp}, the $h = -2$ helicity is obtained by setting $I=J=K=L = 1$. Choosing $\lambda_3^1 = \ket{a}$ and $\lambda_3^2 = \ket{b}$ then, this part of the amplitude reads
\begin{align}
   M^{1111} &= \kappa\frac{[\textbf{12}]^2\braket{\textbf{1}a}^2\braket{\textbf{2}a}^2}{m_h^4}\nonumber\\
   &= \kappa\frac{\braket{a\textbf{2}}^2[\textbf{2}|P_1\ket{a}^2}{m_h^4}\nonumber\\
   &= \kappa\frac{\bra{a}P_2|b]^2}{m_h^2}\,.
\end{align}

Similarly, the choice of $J=2$ and $I=K=L = 1$ yields the spin one contribution, 
\begin{align}\label{key}
   M^{\{1211\}} &= \frac{\kappa[\textbf{12}]^2}{3m_h^4}\left(\braket{\textbf{1}a}^2\braket{\textbf{2}b}   
   \braket{\textbf{2}a} + \braket{\textbf{1}b}\braket{\textbf{2}a}^2\braket{\textbf{1}a}\right) + \frac{\kappa}   
   {4m_h^2}\bra{a}P_1P_2\ket{a}\nonumber\\
   &=		-\frac{\kappa}{3}\left(\frac{\bra{a}P_1|b]}{[ab]^2}\left(\bra{b}P_1|b] - \bra{a}P_1|a]\right)\right) -    
   \frac{\kappa}{4}\bra{a}P_1|b]\nonumber\\
   &= -\frac{\kappa}{12}\bra{a}P_1|b]\left(\frac{4\left(\bra{b}P_1|b] - \bra{a}P_1|a]\right)}{[ab]^2} + 3\right)
   \nonumber\\
   &= \frac{\kappa}{12}\bra{a}P_1|b]\left(\frac{8\left(P_1\cdot (P_a - P_b)\right)}{m_h^2} - 3\right)\,.
\end{align}
$P_b$ can be eliminated through momentum conservation
\begin{equation}\label{kine1}
   P_1\cdot(P_a - P_b) = 2P_1\cdot P_a + \frac12 m_h^2\,.
\end{equation}
Consequently, 
\begin{align}\label{key}
   M^{\{1211\}} = \frac{\kappa}{12}\bra{a}P_1|b]\left(\frac{16P_1\cdot P_a + m_h^2}{m_h^2}\right)\,.
\end{align}
Finally, the scalar mode contribution can be extracted by considering the $I=L=1$ and $J = K = 2$ case
\begin{align}\label{key}
   M^{\left\{1212\right\}} &=  \kappa\frac{[\textbf{12}]^2}{3m_h^4}\left(\braket{\textbf{1}a}^2\braket{\textbf{2}b}^2    
   + \braket{\textbf{1}a}\braket{\textbf{1}b}\braket{\textbf{2}a}\braket{\textbf{2}b} + \braket{\textbf{1}b}   
   ^2\braket{\textbf{2}a}^2\right) + \kappa\frac{[\textbf{12}]\braket{ab}(\braket{\textbf{1}a}\braket{\textbf{2}b} + 
   \braket{\textbf{1}b}\braket{\textbf{2}a})}{2m_h^2} + \kappa\braket{ab}^2\nonumber\\
   &= \kappa\frac{[\textbf{12}]^2\left(\braket{\textbf{1}a}\braket{\textbf{2}b} + \braket{\textbf{1}b}   
   \braket{\textbf{2}a}\right)^2 - [\textbf{12}]^2\braket{\textbf{1}a}\braket{\textbf{1}b}\braket{\textbf{2}a}   
   \braket{\textbf{2}b}}{3m_h^4} + \kappa\frac{[\textbf{12}]\braket{ab}(\braket{\textbf{1}a}\braket{\textbf{2}b} +    
   \braket{\textbf{1}b}\braket{\textbf{2}a})}{2m_h^2} 
   + \frac{\kappa}{2}\braket{ab}^2\nonumber\\
   &= \kappa\frac{\left(\bra{a}P_1|a] - \bra{b}P_1|b]\right)^2}{3m_h^2} - \kappa\frac{\bra{a}P_1P_2\ket{a}\bra{b}   
   P_1P_2\ket{b}}{3m_h^4} + \kappa\frac{\bra{a}P_1|a] - \bra{b}P_1|b]}{2} + \frac{\kappa}{2} m_h^2	\nonumber\\
   &= \kappa\frac{\left(\bra{a}P_1|a] - \bra{b}P_1|b]\right)^2}{3m_h^2} + \kappa\frac{\Trp{P_1P_bP_1P_a}}{3m_h^2}    
   + \kappa\frac{\bra{a}P_1|a] - \bra{b}P_1|b]}{2} + \frac{\kappa}{2} m_h^2\,.
\end{align}
The trace term can be evaluated using \ref{kine1}, the identities
\begin{eqnarray}\label{identities}
   \Trp{\sigma^\mu\overline{\sigma}^\nu\sigma^\rho\overline{\sigma}^\lambda} &= 2(\eta^{\mu\nu}\eta^{\rho\lambda} -    
   \eta^{\mu\rho}\eta^{\nu\lambda} + \eta^{\mu\lambda}\eta^{\rho\nu} + i\epsilon^{\mu\nu\rho\lambda} )
   \nonumber\\
   \Trm{\overline{\sigma}^\mu\sigma^\nu\overline{\sigma}^\rho\sigma^\lambda} &= 2(\eta^{\mu\nu}\eta^{\rho\lambda}    
   - \eta^{\mu\rho}\eta^{\nu\lambda} + \eta^{\mu\lambda}\eta^{\rho\nu} - i\epsilon^{\mu\nu\rho\lambda} )\,,
   \nonumber
\end{eqnarray}
and the fact that $\bra{a}P_1|a] = 2P_1\cdot P_a$ to get, 
\begin{align}M^{\{1212\}} &= \label{key}
   \kappa\frac{\left(\bra{a}P_1|a] - \bra{b}P_1|b]\right)^2}{3m_h^2} 
   + \kappa\frac{4(P_1\cdot P_b) (P_1\cdot P_a) - 2m_\phi^2 m_h^2}{3m_h^2} 
   + \kappa\frac{\bra{a}P_1|a] - \bra{b}P_1|b]}{2} + \frac{\kappa}{2} m_h^2\nonumber\\
   &=\kappa\frac{4\left(2P_1\cdot P_a + \frac12m_h^2\right)^2}{3m_h^2} 
   + \kappa\frac{4(P_1\cdot P_a)(P_1\cdot P_a + \frac12 m_h^2) - 2m_\phi^2 m_h^2}{3m_h^2} 
   + \kappa\frac{4P_1\cdot P_a + m_h^2}{2} + \frac{\kappa}{2} m_h^2\,.
\end{align}
As a second example that we will need shortly, we consider two photons interacting with a massive spin-2 graviton. In this case, the entire amplitude is conveniently determined by the helicities of the massless legs, and is given by \cite{Conde:2016vxs, Arkani-Hamed:2017jhn}
\begin{equation}\label{key}
   M^{h_1,h_2}_{\{\alpha_{1},...,\alpha_{2S}\}} = \frac{g}{m^{2S+h_1+h_2-1 -[g]}}\left(\lambda_1^{S+h_2-   
   h_1}\lambda_2^{S+h_1-h_2}\right)_{\{\alpha_{1},...,\alpha_{2S}\}}[12]^{S+h_1+h_2}\,,
\end{equation}
where $[g]$ is the dimension of the coupling. This is then used to construct the associated 3-point function as
\begin{equation}
   M^{1,-1,\{IJKL\}} = M^{1,-1}_{\{\alpha_{1},...,\alpha_{2S}\}}   
   \lambda_3^{I\alpha_1}\lambda_3^{J\alpha_2}\lambda_3^{K\alpha_3}\lambda_3^{L\alpha_4} =    
   \kappa\frac{\braket{\textbf{3}2}^4}{\braket{12}^2}\,.
\end{equation}
Note here that the only non-zero components of this amplitude are those that correspond to pure spin-2, i.e. those with $I=J=K=L$. This is because we implicitly demand that the powers of un-contracted spinors in a given amplitude be positive, {\it i.e.} $S+h_2-h_1 > 0$ and $S+h_1-h_2 > 0$. This in turn translates into the condition $|h| > S/2$ for a non-vanishing amplitude\footnote{Cases where $h_1\neq -h_2$ results in the final amplitude being zero as a result of Bose symmetry}. In field theory, this is equivalent to the statement that the spin-1 contribution is automatically zero due to the Landau-Yang theorem\footnote{By way of self-containedness, we recall here that the Landau-Yang theorem essentially says that, on-shell, a massive spin-1 particle cannot decay into two photons.}, and that the photon does not couple to the scalar mode of the graviton in pure gravity since its stress-energy tensor is traceless. This is exactly what we found was the source of the discontinuity in the more familiar Lagrangian formulation with the introduction of St\"uckleberg fields.

\section{Four-point functions from BCFW}
Having derived the individual helicity components of the 3pt amplitudes, 4-point amplitudes can now be computed from BCFW relations \cite{Britto:2004ap, Britto:2005fq}, using the formula
\begin{equation}
   A_n = i\sum_{z_{ib}}\sum_{h}A_L(z_{ib})\frac{1}{P_{ib}^2-m_{ib}^2}A_R(z_{ib}),\label{eq:recrelmass}\,.
\end{equation}
In order to see the vDVZ discontinuity, we will need to calculate two sets of amplitudes: one that couples to the scalar mode of the graviton and one that does not. We choose to calculate the photon-scalar amplitude in one case, and the scalar-scalar amplitude in the other, with both mediated by a massive graviton.
\subsection{Photon-Scalar Amplitude}
Following \cite{Burger:2017yod}, we shift momenta $2,3$ and consider the diagram
\begin{equation}\label{fdiags1}
\begin{gathered}
\begin{fmfgraph*}(120,80)
\fmfleft{i1,i2}
\fmfright{o1,o2}
\fmf{photon,label=$1^{\mp1}$,label.side=left}{i1,v1}
\fmf{plain,label=$4^0$,label.side=left}{i2,v1}
\fmf{photon,label=$2^{\pm1}$,label.side=left}{v1,o1}
\fmf{plain,label=$3^0$,label.side=left}{v1,o2}
\fmfblob{.30w}{v1}
\end{fmfgraph*}
\end{gathered}
~~~=~~~\mathlarger{\mathlarger{\mathlarger{\mathlarger{‎‎\sum}}}}_{s}
\begin{gathered}
\begin{fmfgraph*}(120,80)
\fmfleft{i1,i2}
\fmfright{o1,o2}
\fmf{photon,label=$1^{\mp1}$,label.side=left}{i1,v1}
\fmf{fermion,label=$4^0$,label.side=left}{i2,v2}
\fmf{dbl_wiggly}{v1,v2}
\fmf{photon,label=$2^{\pm1}$,label.side=left}{v1,o1}
\fmf{fermion,label=$3^0$,label.side=left}{v2,o2}
\fmflabel{$_{-s}$}{v1}
\fmflabel{$_{s}$}{v2}
\end{fmfgraph*}
\end{gathered}
\end{equation}
Now from eq. \ref{eq:recrelmass}, there are five terms that can contribute to the amplitude
\begin{align}\label{key}
   A_4[1^+,2^-,3,4] &= A[1^+,\hat{2}^-,\hat{P}_h^{-2}]\frac{1}{P_h^2 - m_h^2}A[-\hat{P}_h^{+2},\hat{3},   
   4]\nonumber \\&~~~~~+ A[1^+,\hat{2}^-,\hat{P}_h^{+2}]\frac{1}{P_h^2 - m_h^2}A[-\hat{P}_h^{-2},\hat{3},4]
   \nonumber\\&~~~~~+ A[1^+,\hat{2}^-,\hat{P}_h^{-1}]\frac{1}{P_h^2 - m_h^2}A[-\hat{P}_h^{+1},\hat{3},4]
   \nonumber\\&~~~~~+ A[1^+,\hat{2}^-,\hat{P}_h^{+1}]\frac{1}{P_h^2 - m_h^2}A[-\hat{P}_h^{-1},\hat{3},4]
   \nonumber\\&~~~~~+    
   A[1^+,\hat{2}^-,\hat{P}_h^{0}]\frac{1}{P_h^2 - m_h^2}A[-\hat{P}_h^{0},\hat{3},4]\nonumber\\
   &= A[1^+,\hat{2}^-,\hat{P}_h^{-2}]\frac{1}{P_h^2 - m_h^2}A[-\hat{P}_h^{+2},\hat{3},4]\,,
\end{align}
with the opposite helicity 3-point function determined by complex conjugation. As a result, the full amplitude is
\begin{eqnarray}
   A_4 = \left(\kappa\frac{\braket{2a}^4}{\braket{12}^2}\right)\times \frac{1}{P_h^2 -    
   m_h^2}\times\left( \kappa\frac{\bra{b}P_3|a]^2}{m_h^2}\right)\,.
\end{eqnarray}
In this form\footnote{The positive helicity-2 graviton piece does not contribute. See, for example, eq.4.19 of \cite{Burger:2017yod} for a detailed discussion of this point.}, we can recover the massless amplitude by taking $m_h \longrightarrow 0$ and $a\longrightarrow P_h$. Using the fact\footnote{Using momentum conservation $-P_h = P_1 + P_2$, we can use the spinor representation to write this as $|1]\bra{1} + |2]\bra{2} = -|a]\bra{a} - |b]\bra{b}$, then contract with $[1|$ from the left and $\ket{a}$ from the right.} that $[12]\braket{2a} = [1b]\braket{ab} = m_h[1b]$, we find
\begin{eqnarray}
   A_4 &=& \left(\kappa\frac{\braket{2a}^4}{\braket{12}^2}\right)\times \frac{1}{P_h^2 -    
   m_h^2}\times\left( \kappa\frac{[1|P_bP_3|a]^2}{[1b]^2\braket{ab}^2}\right)\nonumber\\
   &=& \left(\kappa\frac{\braket{2a}^4}{\braket{12}^2}\right)\times \frac{1}
   {P_h^2 - m_h^2}\times\left( \kappa\frac{[1|   
   (P_a + P_2)P_3|a]^2}{[1b]^2m_h^2}\right)\nonumber\\
   &=& \left(\kappa\frac{\braket{2a}^4}{\braket{12}^2}\right)\times \frac{1}{P_h^2 -    
   m_h^2}\times\left( \kappa\frac{[12]^2\bra{2}P_3|a]^2}{[1b]^2m_h^2}\right)\nonumber\\
   &\stackrel{m_{h}\to 0}{\longrightarrow}&
   \frac{\kappa^2}{P_h^2}\frac{\braket{2P_h}^2\bra{2}P_3|P_h]^2}{\braket{12}^2}\nonumber\\
   &=& \kappa^2\frac{\bra{2}P_3|1]^2}{P_h^2}\,.
\end{eqnarray}
In order to have this agree with eq. 4.25 of $\cite{Burger:2017yod}$, we rescale 
$\kappa \longrightarrow \tilde{\kappa} = \frac{\kappa}{2}$ to find
\begin{equation}\label{lbeqn}
   A_4 = \frac{\tilde{\kappa}^2}{4}\frac{\bra{2}P_3|1]^2}{P_h^2}\,,
\end{equation}
in the $m_{h}\to 0$ limit. This rescaling of the coupling corresponds to the choice of normalisation used in the Einstein-Hilbert action.	

\subsection{Scalar-Scalar Amplitude}

Next, we consider the $2\longrightarrow 2$ scattering of massive scalars in massive gravity. For simplicity, we will take  both scalars to have the same mass $m_\phi$ and the massive graviton to again have mass $m_h$.
Again, we will compute the 4-point amplitude using BCFW with momenta 1 and 3 shifted, as coded in the diagram
\begin{equation}\label{fdiags1}
\begin{gathered}
\begin{fmfgraph*}(120,80)
\fmfleft{i1,i2}
\fmfright{o1,o2}
\fmf{plain,label=$1$,label.side=left}{i1,v1}
\fmf{plain,label=$4$,label.side=left}{i2,v1}
\fmf{plain,label=$2$,label.side=left}{v1,o1}
\fmf{plain,label=$3$,label.side=left}{v1,o2}
\fmfblob{.30w}{v1}
\end{fmfgraph*}
\end{gathered}
~~~=~~~\mathlarger{\mathlarger{\mathlarger{\mathlarger{‎‎\sum}}}}_{s}
\begin{gathered}
\begin{fmfgraph*}(120,80)
\fmfleft{i1,i2}
\fmfright{o1,o2}
\fmf{plain,label=$1$,label.side=left}{i1,v1}
\fmf{plain,label=$4$,label.side=left}{i2,v2}
\fmf{dbl_wiggly}{v1,v2}
\fmf{plain,label=$2$,label.side=left}{v1,o1}
\fmf{plain,label=$3$,label.side=left}{v2,o2}
\fmflabel{$_{-s}$}{v1}
\fmflabel{$_{s}$}{v2}
\end{fmfgraph*}
\end{gathered}
\end{equation}
To decompose the massive shifted lines into massless ones, we write
\begin{equation}\label{key}
   P_1 = k_1 + xk_3,~~~~~~P_3 = k_3 + xk_1\,,
\end{equation}
where $k_i^2 = 0$ and $x = -\frac{m_\phi^2}{2k_1\cdot k_3}$. With this in mind, it is clear that the on-shell condition $P_i^2 = m_\phi^2$ holds. Subsequently, the vectors are continued to complex values by writing
\begin{equation}\label{key}
   \hat{P_1}(z) = P_1 + z\eta,~~~~~\hat{P_3}(z) = P_3 - z\eta\,,
\end{equation}
with $\eta\cdot P_1 = \eta\cdot P_3 = 0$ and $\eta^2 = 0$. An obvious choice, given our decomposition, is 
$\eta = \ket{1}[3|$. Again, there are five terms that contribute to the amplitude in \eqref{eq:recrelmass}, 
\begin{eqnarray}\label{key}
   A_4[1,2,3,4] &= A[1,\hat{2},\hat{P}_h^{-2}]\frac{1}{P_h^2 - m_h^2}A[-\hat{P}_h^{+2},\hat{3},4] 
   \nonumber\\
   &+ A[1,\hat{2},\hat{P}_h^{+2}]\frac{1}{P_h^2 - m_h^2}A[-\hat{P}_h^{-2},\hat{3},4]
   \nonumber\\&+ A[1,\hat{2},\hat{P}_h^{-1}]\frac{1}{P_h^2 - m_h^2}A[-\hat{P}_h^{+1},\hat{3},4]
   \nonumber\\&+ A[1,\hat{2},\hat{P}_h^{+1}]\frac{1}{P_h^2 - m_h^2}A[-\hat{P}_h^{-1},\hat{3},4]
   \nonumber\\&+ A[1,\hat{2},\hat{P}_h^{0}]\frac{1}{P_h^2 - m_h^2}A[-\hat{P}_h^{0},\hat{3},4]\,,
   \label{scalar-mode}
\end{eqnarray}
and we will evaluate each contribution separately.

\subsubsection{Scalar Graviton Contribution}
This is the last term in \eqref{scalar-mode}, the piece contributed by the scalar mode of the graviton. This takes the 
form
\begin{eqnarray}
   &&\kappa^2\left(\frac23\right)\left[\frac{4\left(2P_1\cdot P_a + \frac12m_h^2\right)^2}{2m_h^2} 
   + \frac{4(P_1\cdot P_a)(P_1\cdot P_a + \frac12 m_h^2) - 2m_\phi^2 m_h^2}{2m_h^2} + \frac{3(4P_1\cdot P_a 
   + m_h^2)}{4} + \frac{3}{4} m_h^2\right]\nonumber\\
   &\times&\frac{1}{P_h^2 - m_h^2}\left[\frac{4\left(2P_1\cdot P_a + \frac12m_h^2\right)^2}{2m_h^2} 
   + \kappa\frac{4(P_1\cdot P_a)(P_1\cdot P_a + \frac12 m_h^2) - 2m_\phi^2 m_h^2}{2m_h^2} 
   + \frac{3(4P_1\cdot P_a + m_h^2)}{4} + \frac{3}{4} m_h^2\right]\nonumber\,.
\end{eqnarray}
To understand the $m_h\longrightarrow 0$ limit of this amplitude, we first note that
\begin{equation}
   \hat{P_1}\cdot P_h = -(\hat{P_1}\cdot \hat{P_1} + \hat{P_1}\cdot \hat{P_2}) = -(m_\phi^2 + \hat{P_1}\cdot    
   \hat{P_2}) = -(m_\phi^2 +\frac12m_h^2 - m_\phi^2) = -\frac{m_h^2}{2}\,,
\end{equation}
so that $P_{1}\cdot P_{h}\to 0$ as $m_{h}$ is taken to zero. Consequently,
\begin{equation}
   A^0_4[1,2,3,4] = \frac43\kappa^2\frac{m_\phi^4}{t} = \frac13\tilde{\kappa}^2\frac{m_\phi^4}{t}\,,
\end{equation}
where, as usual, we have defined $t = (P_1 + P_2)^2$. This also has virtue of having the correct mass dimension and vanishes in the $m_\phi = 0$ limit, as expected.

\subsubsection{Spin-1 Graviton Contribution}
The spin-1 contribution to the full amplitude is 
\begin{align}\label{key}
   A^{(1)}[1,2,3,4] &=
   \frac{\kappa^2}{144}\bra{a}P_1|b]\left(\frac{16P_1\cdot P_a + m_h^2}{m_h^2}\right)\times\frac{1}{P_h^2 +    
   m_h^2}\times\bra{b}P_3|a]\left(\frac{16P_3\cdot P_a + m_h^2}{m_h^2}\right)\nonumber\\
   &= \frac{\kappa^2}{144(P_h^2 + m_h^2)}\bra{a}P_1|b]\bra{b}P_3|a]\left(\frac{16P_1\cdot P_a + m_h^2}   
   {m_h^2}\right)\left(\frac{16P_3\cdot P_a + m_h^2}{m_h^2}\right)\,.
 \end{align}
 To find its massless limit, we again take $P_a \longrightarrow P_h$, $\ket{b} \longrightarrow m_h\ket{\overline{b}}$,
 and $|b] \longrightarrow m_h|\overline{b}]$ to find
 \begin{align}\label{key}
    A^{(1)}_{m_h\rightarrow 0}[1,2,3,4] &= \frac{\kappa^2 m_h^2}{144(P_h^2 + m_h^2)}  
    \bra{P_h}P_1|\overline{b}]\bra{\overline{b}}P_3|P_h]\left(\frac{16P_1\cdot P_h + m_h^2}{m_h^2}\right)   
    \left(\frac{16P_3\cdot P_h + m_h^2}{m_h^2}\right)\nonumber\\
    &= \frac{49\kappa^2 m_h^2}{144(P_h^2 + m_h^2)}\bra{P_h}P_1|\overline{b}]   
    \bra{\overline{b}}P_3|P_h]\,\,\stackrel{m_{h}\to 0}{\longrightarrow} 0\,,
 \end{align}
 as anticipated.

\subsubsection{Spin-2 Graviton Contribution}
Finally, we evaluate the spin-2 part of the amplitude, given by
\begin{eqnarray}\label{key}
   A^{(2)}[1,2,3,4] &=& A^{-+}[1,2,3,4] + A^{+-}[1,2,3,4]\nonumber\\
   &=& A[1,\hat{2},\hat{P}_h^{-2}]\frac{1}{P_h^2 - m_h^2}A[-\hat{P}_h^{+2},\hat{3},4] + A[1,\hat{2},\hat{P}   
   _h^{+2}]\frac{1}{P_h^2 - m_h^2}A[-\hat{P}_h^{-2},\hat{3},4]\nonumber\\
   &=& \left( \kappa\frac{\bra{b}P_1|a]^2}{m_h^2}\right)\times\frac{1}{P_h^2 
   - m_h^2}\times\left( \kappa\frac{\bra{a}P_3|b]^2}{m_h^2}\right) + a\leftrightarrow b\,,
\end{eqnarray}
where in the last line we have used the 3-point function calculated earlier. Following the same $m_h \longrightarrow 0$ procedure as in the spin zero case we find, with some algebra, that the amplitude ultimately reduces to
\begin{equation}\label{key}
A^{(2)}[1,2,3,4] = \frac{\kappa^2}{t}\left(\frac{\Trp{P_1P_aP_3P_b}\Trm{P_1P_aP_3P_b}}{m_h^2} + \frac{\Trp{P_3P_aP_1P_b}\Trm{P_3P_aP_1P_b}}{m_h^2}\right)\,,
\end{equation}
where, for example, $\Trpm{P_1P_aP_3P_b} = 2(P_1\cdot P_3)(P_a\cdot P_b) \pm 2i\epsilon_{\mu\nu\rho\sigma}P_{1}^\mu P_{a}^\nu P_3^\rho P_b^\sigma\,$.
The antisymmetric piece can be evaluated by noting that $\epsilon_{\mu\nu\rho\sigma}P_{1}^\mu P_{a}^\nu P_3^\rho P_b^\sigma = \sqrt{\det(G)}$, where $G$ is the {\it Gram matrix}, whose determinant
\begin{align*}\label{key}
\det(G) &= \left|\begin{matrix}
P_1^2 & P_1\cdot P_a & P_1\cdot P_3 & P_1\cdot P_b\\ 
P_1\cdot P_h & P_a^2 & P_a\cdot P_3 & P_a\cdot P_b \\ 
~~P_1\cdot P_3~~ & ~~P_a\cdot P_3~~ & ~~P_3^2~~ & ~~P_3\cdot P_b~~ \\ 
P_1\cdot P_b & P_a\cdot P_b & P_3\cdot P_b& P_b^2
\end{matrix}\right| = \left|\begin{matrix}
m_\phi^2 & P_1\cdot P_a & P_1\cdot P_3 & 0 \\ 
P_1\cdot P_a & 0 & P_a\cdot P_3 & \frac12 m_h^2 \\ 
P_1\cdot P_3 & P_a\cdot P_3 & m_\phi^2 & 0 \\ 
0 & \frac12 m_h^2 & 0 & 0
\end{matrix}\right|\\\\
&= m_h^4(-m_\phi^4 + (P_1\cdot P_3)^2)\\
&= \frac14 m_h^4\left((s-2m_\phi^2)^2 - 4m_\phi^4\right)\,.
\end{align*} 
Substituting back into the trace gives
\begin{equation}\label{key}
   \Trpm{P_1P_aP_3P_b} = 2m_h^2\left((P_1\cdot P_3) \pm \frac i2\sqrt{(s-2m_\phi^2)^2 - 4m_\phi^4}\right)\,.
\end{equation}
The second trace term is evaluated analogously and found to be the same, so that the final expression for the amplitude is \cite{Guevara:2017csg, Cachazo:2017jef}
\begin{equation}\label{key}
   A^{(2)}[1,2,3,4] = 2\kappa^2\frac{(s + 2m_\phi^2)^2 - 2m_\phi^4}{t} = \frac12\tilde{\kappa}^2\frac{(s 
   + 2m_\phi^2)^2 - 2m_\phi^4}{t}\,.
\end{equation}
Pulling this all together gives the full amplitude 
\begin{equation}\label{key}
   A[1,2,3,4] = A^{0}[1,2,3,4] + A^{(1)}[1,2,3,4] + A^{(2)}[1,2,3,4]\,,
\end{equation}
whose massless limit is
\begin{align}\label{key}
   A_{m_h\rightarrow 0}[1,2,3,4] &= A^{0}_{m_h\rightarrow 0}[1,2,3,4] + A^{(2)}_{m_h\rightarrow 0}[1,2,3,4]
   = \frac{\tilde{\kappa}^2}{2t}\left[(s + 2m_\phi^2)^2 - 2m_\phi^4 - \frac23 m_\phi^4\right]\,.
\end{align}
The classical Newtonian potential (in fourier space) is recovered by first going to the center-of-mass frame where, 
\begin{equation}\label{key}
   t = -\vec{q}^{\,\,2},~~~~~s + 2m_\phi^2 = 2m_\phi^2 + 4m_\phi^2\vec{p}^{\,\,2} + \cdots
\end{equation}
If, in addition, we take $\tilde{\kappa}^2 = 32\pi G$, the amplitude becomes
\begin{equation}\label{key}
   A_{m_h\rightarrow 0}^{COM}[1,2,3,4] = \frac{16\pi G}{\vec{q}^{\,\,2}}\left[\frac{4}{3}m_\phi^4 + 16\vec{p}^{\,\,2}\,m_   
   \phi^4 + \cl{O}(\vec{p}^{\,\,4}) \right]\,.
\end{equation}
To find the classical potential, we use the relation \cite{Feinberg:1988yw}
\begin{equation}\label{key}
T_{fi}^{COM} = \frac{A_{fi}^{COM}}{4E^2}\,,
\end{equation}
where, if we write $E = m_\phi + \frac{\vec{p}^{\,\,2}}{2m_\phi} + \cdots$and take $m_\phi \gg \vec{p}^{\,\,2}$,
\begin{equation}
   T_{fi}^{COM} =  \frac{16\pi Gm_\phi^{2}}{\vec{q}^{\,\,2}}\left[\frac{1}{3} + 4\vec{p}^{\,\,2} 
   + \cl{O}(\vec{p}^{\,\,4}, m_{\phi}^{2}) \right]\,.
\end{equation}
To first order, this is exactly $\frac{4}{3}$ the classical Newtonian potential (in momentum space):
\begin{equation}\label{key}
   T_{fi}^{COM}(0) = \frac{4}{3}\left(\frac{4\pi Gm_\phi^2}{\vec{q}^{\,\,2}}\right)\,.
\end{equation}
Then, by the standard arguments, these expressions can be reconciled by rescaling the coupling 
$G\longrightarrow \frac{3}{4} G$. Of course, this comes with the cost that the light bending angle derived 
from \eqref{lbeqn} will now be $3/4$ of that predicted by general relativity, manifesting the vDVZ discontinuity.

\section{Conclusions \& Future Work}
While completing this article, the LIGO/VIRGO collaboration made the first multi-messenger detection of a binary inspiralling neutron star system \cite{PhysRevLett.119.161101}. With a constraint on the fractional difference in speed between the emitted gravitational wave and the speed of light of 
\begin{eqnarray}
   -3\times 10^{-15}\leq \frac{\Delta v}{c} \leq +7\times 10^{-16}\,,
\end{eqnarray}
this remarkable observation establishes the most stringent constraint on the mass of a graviton to date. For all intents and purposes, the graviton is massless. We don't disagree. However, giving the graviton a small, but nonvanishing mass $m_{h}$ has always provided a useful regulator in the study of the field-theoretic properties of gravity. This continues to be true today. Toward this end, the massive gravity that this note concerns itself with furnishes a useful laboratory to explore some of the latest developments in ``on-shell" quantum field theory. In particular, we have focussed on trying to understand the famous vDVZ discontinuity that arises when the regulator $m_{h}\to 0$.
We have shown that the vDVZ discontinuity exists at the level of the on-shell amplitudes \textit{without} the unnecessary baggage of an off-shell action, gauge symmetries or polarization. Indeed, all that was required was a little knowledge of the Poincar\'e group, dimensional analysis and Newtonian gravity. We have, however, only worked at linear order, and one interesting future direction would be to consider higher order amplitudes (using BCFW recursion) to see if and how the Veinshtein mechanism appears to suppress contributions from the discontinuity. Additionally, there exists some controversy in the literature as to whether or not the vDVZ discontinuity persists in Einstein spaces satisfying $R_{\mu\nu} = \Lambda g_{\mu\nu}$ with non-vanshing cosmological constant $\Lambda$ and $m_{h}^{2}/\Lambda\to 0$ (see, for example, \cite{Dilkes:2001av} and references therin). In principle, an on-shell analysis, unencumbered by the usual subtlties of Lagrangians, gauge-fixing and the like, should provide a cleaner resolution of this curious issue. We leave these ideas for future work. 
 
\section*{Acknowledgements}
We are grateful to Nima Arkani-Hamed, Daniel Burger, Raul Carballo-Rubio and Bryan Gaensler for useful discussion at various stages of this work. NM is supported by the South African Research Chairs Initiative of the Department of Science and Technology and the National Research Foundation of South Africa. Any opinion, finding and conclusion or recommendation expressed in this material is that of the authors and the NRF does not accept any liability in this regard.
\end{fmffile}
	\nocite{*}
	\bibliographystyle{rubaton}
	\bibliography{mainbib}
\end{document}